\documentclass{appolb}
\usepackage{epsfig}

\begin{document}
\title{ What Do We Expect to Learn from\\ Very High Energy Cosmic Ray
Observations?%
\thanks{Presented at the XVII Mazurian Lakes School of Physics "Growth
    Points of Nuclear  Physics  A.D. 2001", September 2--9, 2001 Krzyze, Poland}%
}
\author{H. Rebel
\address{Forschungszentrum Karlsruhe, Institut f\"ur Kernphysik}
}
\maketitle
\begin{abstract}

By a short tour through the exciting field of very high-energy
and ultra high-energy cosmic rays studies, a brief review is given about the
current questions approached, in particular by the KASCADE experiment and
the Pierre Auger project. The present status of the investigations of the
knee region of the cosmic ray spectrum by KASCADE is presented and open
problems are discussed.

\end{abstract}

\PACS{96.40}

\section{Introduction}

Many kinds of radiation exist in the Universe, electromagnetic radiation and
particles with a broad range of wavelengths and energies, respectively. Some
of the radiation is produced in stars and galaxies, while some is the
cosmological background radiation, a relic from the cosmic evolution. Among
this radiation, the most energetic are cosmic rays particles, dominantly
protons, helium, carbon, nitrogen up to iron ions in a appreciable amount.

They continuously bombard our Earth from the cosmos by an isotropic stream
of high energy particles. These cosmic rays were discovered in 1912 by the
Austrian Victor Hess [1] through a series of balloon flights, in which he
carried electrometers to over 5000\,m altitudes. Nowadays we know that the
energy spectrum of these particles extends from 1\,GeV to beyond $10^{20}$\,eV
(100\,EeV), to the highest energies of known individual particles in the
Universe. However, we have only a rudimentary understanding, where these
particles are coming from, how they are accelerated to such high energies
and how they propagate through the interstellar space. The difficulty is
that cosmic rays are overwhelmingly charged particles (stripped nuclei),
and the galactic magnetic fields are sufficiently strong to scramble their
paths. Perhaps except at highest energies cosmic rays have lost all their
memory about the location of the emission sources, when they eventually
arrive the Earth's atmosphere. Hence the only observable quantities, which
may give us some information are \textit{the energy distribution} and \textit{the elemental
composition} of primary cosmic rays, at highest energies eventually with
\textit{deviations from isotropic incidence}. The experimental determination are
topics of contemporary research, especially in regions which exceed
the energies provided by man-made accelerators.

\vbox{
\epsfig{file=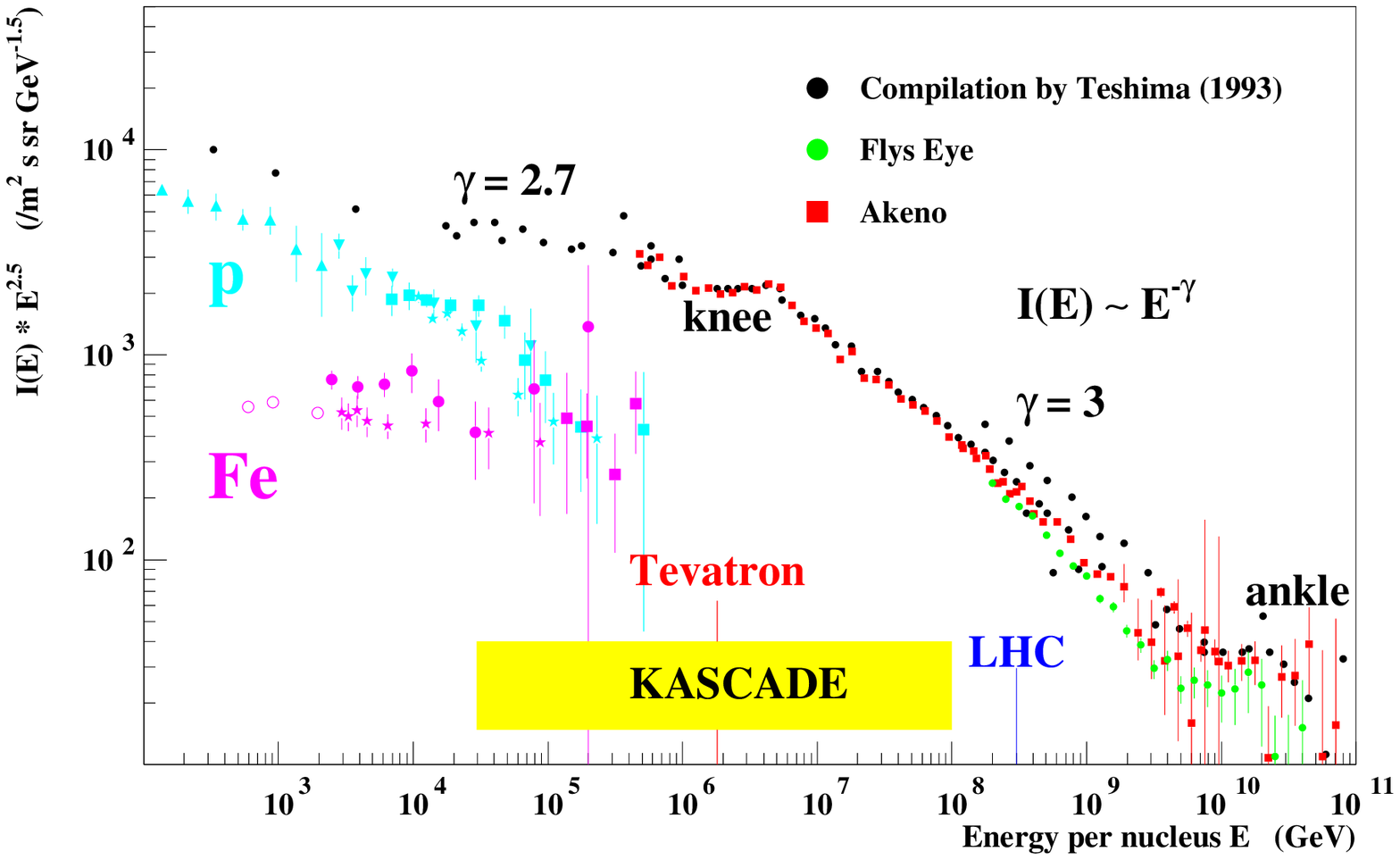,clip=,width=12cm}

\centerline{\em Fig.1 Primary energy spectrum of cosmic rays}

}

The following brief overview will give an impression about the current
experimental activities with their astrophysical motivations.

The rather featureless energy spectrum of primary cosmic rays (Fig.1) comprises more
than 12 orders of magnitude in the energy scale. It follows an overall
power-law ($\propto \mbox{E}^{-2.7}$: Note that the flux is multiplied by E$^{2.7}$) with a
distinct change around $10^{15}$\,eV, called the \textit{"knee"}. This feature, still not
consistently explained, has been discovered 40 years ago by German~Kulikov
and George~Khristiansen from the Moscow~State~University [2] with studies of
the intensity spectrum of Extensive Air Showers (EAS), of the so-called
shower size, which roughly reflects the primary energy. The flux of primary
cosmic rays falls from 1\,$\mbox{particle}/\mbox{m}^2\cdot \mbox{s}$ to 
1\,$\mbox{particle}/\mbox{km}^2\cdot \mbox{century}$
at highest energies. A great deal of interest and
current efforts concern the shape of the spectrum in the EeV-region, above
$10^{18}$\,eV, where the spectrum seems to flatten (\textit{"ankle"}), especially around
$5 \cdot 10^{19}$\,eV, with the theoretically predicted Greisen-Zazepin-Kuzmin
cut-off [3], due to the photo-interaction of protons with the
2.7\,K-background
radiation. The AGASA experiment in Akeno (Japan) [4], in
particular, has shown that this limit does not exist, and this fact is an
issue of extreme astrophysical and cosmological relevance, establishing an
enigma.

Below $10^{14}$\,eV the flux of particles is sufficiently large that individual
nuclei can be studied by flying detectors in balloons and satellites. From
such \textit{direct} experiments we know that the majority of particles are nuclei of
common elements. Around 1\,GeV the abundances are strikingly similar to those
found in ordinary material of the solar system. Striking exceptions are the
abundance of elements like Li, Be, and B, overabundant since originating
from spallation of heavier nuclei in the interstellar medium.

\section{Methodical features and techniques}

Above $10^{14}$\,eV the techniques used to study cosmic rays employ the phenomenon
of Extensive Air Showers discovered independently by Auger et al. [5] and
Kohlh\"orster et al. [6] in 1938.

\vbox{
\vspace{0.5em}
\epsfig{file=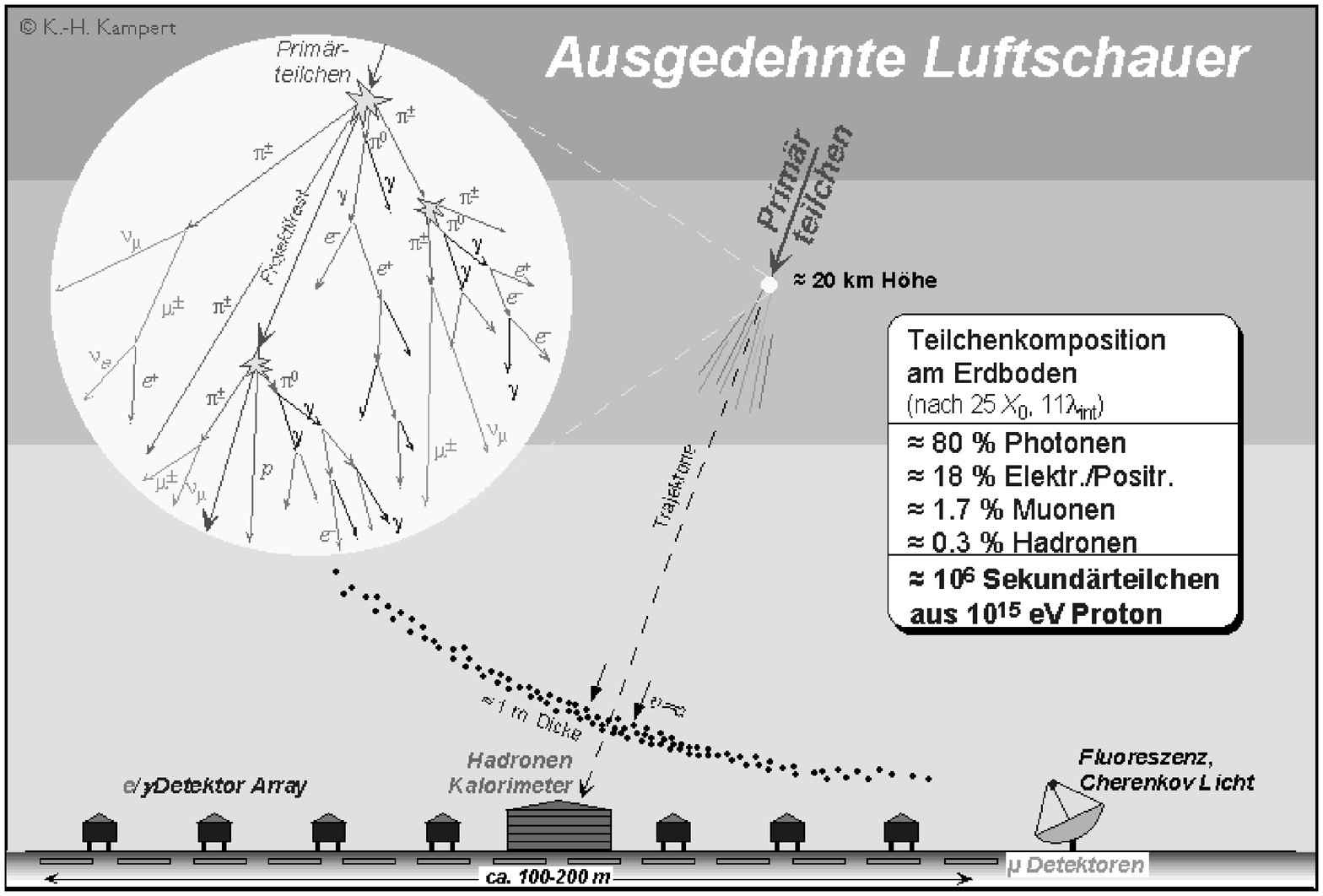,clip=,angle=-90,width=12cm}

\vspace{0.3em}

\centerline{\em Fig.2 Extensive air showers --- Ausgedehnte Luftschauer ---
  Grandes gerbes}

\vspace{0.3em}
}

Most of the produced particles in the hadronic interactions are pions and
kaons, which can decay into muons and neutrinos before interacting, thus
producing the most penetrating component of atmospheric showers. The most
intensive component - electrons and photons - originates from the fast
decay of neutral pions into photons, which initiate electromagnetic showers,
thus distributing the originally high energy over millions of charged
particles. The backbone of an air shower is the hadronic component of
nucleons, pions and more exotic particles (Fig.2).

The electromagnetic component is accompanied by an additional EAS
phenomenon, the production of atmospheric Cerenkov light which carries
further information about the shower development. For EAS with higher
energies ($> 10^{17}$\,eV) also the nitrogen fluorescence light induced in the
atmosphere can be observed.

However, in ground -based experiments, in general, we are not in the
situation to see the longitudinal development, we observe only the developed
status of the air shower cascade at a certain observation level. From the
observables, that means from \textit{the intensity, the lateral and eventually the
energy distributions}, we have to deduce the properties of the primary
particle.

The intensity and the width of the lateral distributions of the three
components are very different. The muons, for example, extend to several
hundred meters as most of them are produced very high in the atmosphere.
Therefore, even a small transverse momentum imparted to them in the
production can lead to large distances from the shower axis.

In an EAS experiment the lateral distributions of the particles are sampled
by more or less regular arrangements of a large number of detectors which
cover only a small fraction of the total area. This sampling is an
additional source of fluctuations which add to the large spread resulting
from the inherent statistical fluctuations of the shower development in the
atmosphere. As an example the photo shows the KASCADE [7] detector
arrangement, installed in Forschungszentrum~Karlsruhe.

{\sloppy
KASCADE (Fig.3) is a multi-component detector array: a field of electron-muon
counters and a central detector set up, which is a complex arrangement of
several types of detectors, basically a iron sampling calorimeter for hadron
measurements and multiwire proportional chambers below, for studies of the
higher energy muon component, and other detectors for special purposes.

}

\vspace{0.3em}

\vbox{

\centerline{\epsfig{file=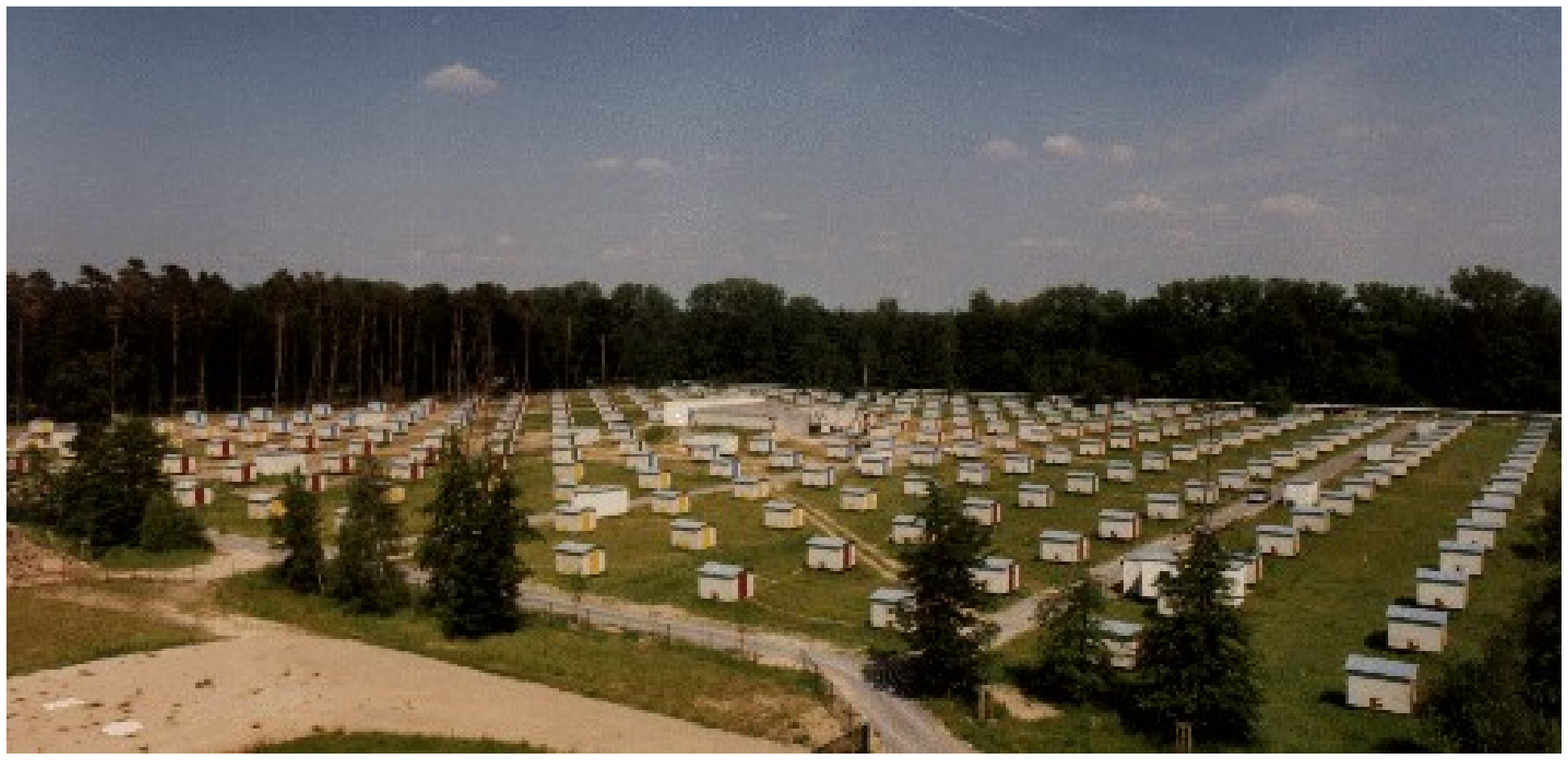,clip=,width=9.cm}}

\setbox0=\hbox{\em Fi g.3} 
{\em\hangafter1\hangindent\wd0 \noindent
Fig.3 KASCADE detector array with a field array, a central detector for
measuring the hadron component and the muon component at various energy
thresholds. In addition there is a muon tracking detector arrangement in a
tunnel.

}

}

I would like to mention that the KASCADE collaboration is just extending the
detector, together with the University of Torino, in order to register
efficiently showers at energies up to beyond the LHC energy: KASCADE~GRANDE
[8] distributes the detector stations over an $800\times 700\,m^2$ area.

\vbox to 0pt{
\vspace{0.3em}
\noindent{\epsfig{file=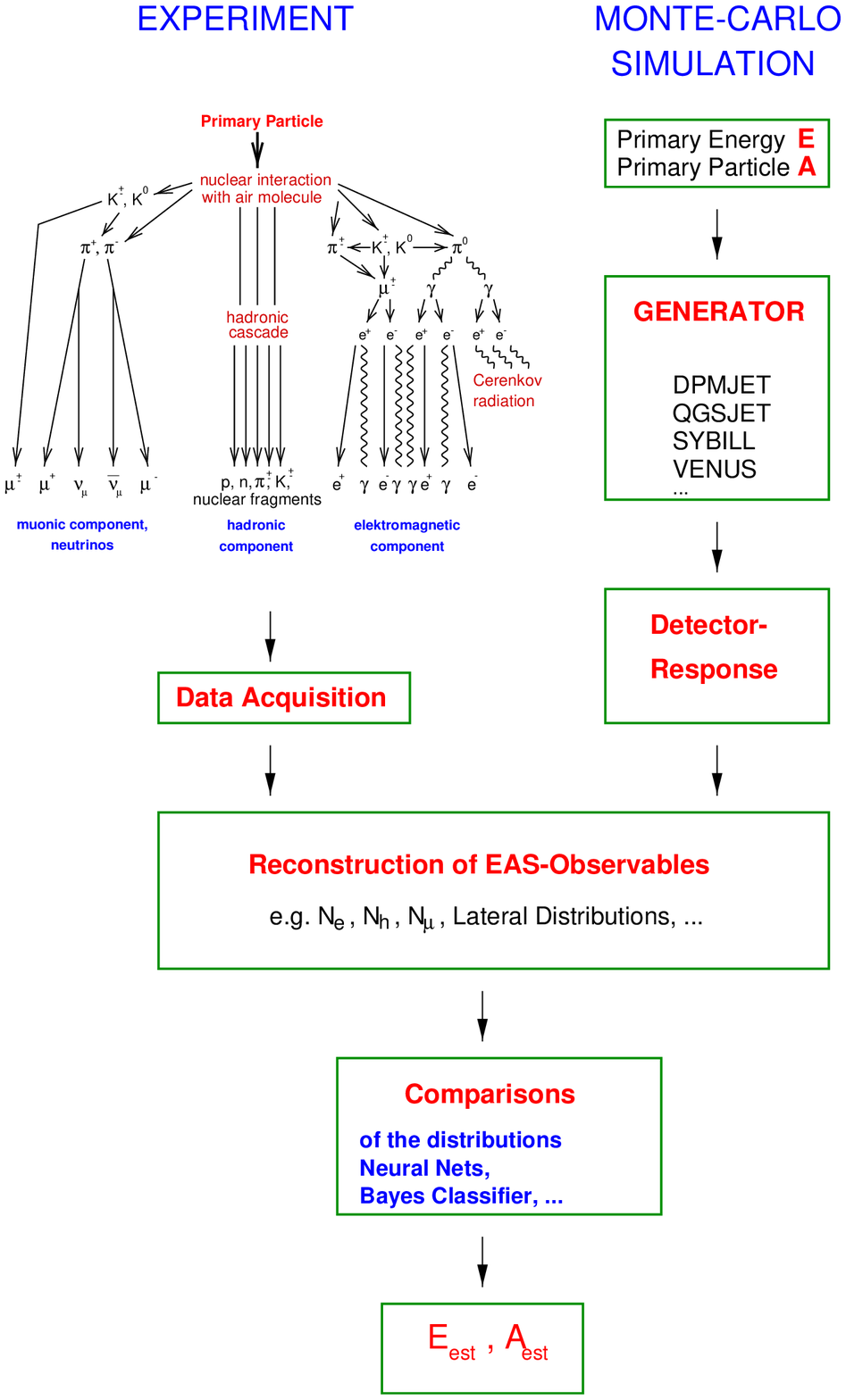,clip=,height=11.3cm}}

\noindent\em \hbox{
  \vtop{
    \hbox{Fig.4}
  }
  \vtop{
    \hbox{General scheme of the analysis}
    \hbox{of EAS observations}
  }
}
\vss
}

{\em\sloppy
\hangafter=-99\hangindent=7cm
The general scheme of inference in a modern EAS experiment is displayed in
Fig.4, indicating also the involved difficulties.

\hangafter=-24\hangindent=7cm
The identification of differences in EAS which result from differences in
mass of the primary particle requires a modelling of shower development in
the atmosphere. For that Monte~Carlo programs of the EAS development like
the Karlsruhe CORSIKA program have been developed [9]. It is under continuous
modification and improvement. A prerequisite for the Monte~Carlo procedures
is a knowledge about particle production in high-energy hadronic
interactions. Since the energy region of our interest exceeds the particle
energies provided at man made accelerators, we rely on model descriptions
which extend the present knowledge to a "terra incognita", on basis of more
or less detailed theoretical approaches of phenomenological nature and with
QCD inspired ideas. (The development of such models is an item of its own).
The CORSIKA code includes various models, presently en vogue as options, and
in fact, the model dependence is an obvious feature in the actual
comparisons with the experimental data.

}

A multi-detector experiment observing simultaneously all major EAS
components with many observables provides some possibilities to test the
hadronic interaction models and to specify the most consistent one.

The stochastic character of the huge number of cascading interactions in the
shower development implies considerable fluctuations of the experimentally
observed EAS parameters and of the corresponding simulated showers as well,
clouding the properties of the original particle. The inherent (unavoidable)
fluctuations establish an important and intriguing difficulty of the EAS
analysis and need adequate response of the analysis methods.

The further processing is to compare real data with pseudo experimental data
on equal level, including the detector response and expressed by various
reconstructed shower variables: \textit{shower intensity, the lateral, arrival time
and eventually energy distributions of the various EAS components.}

\vbox{
\centerline{\epsfig{file=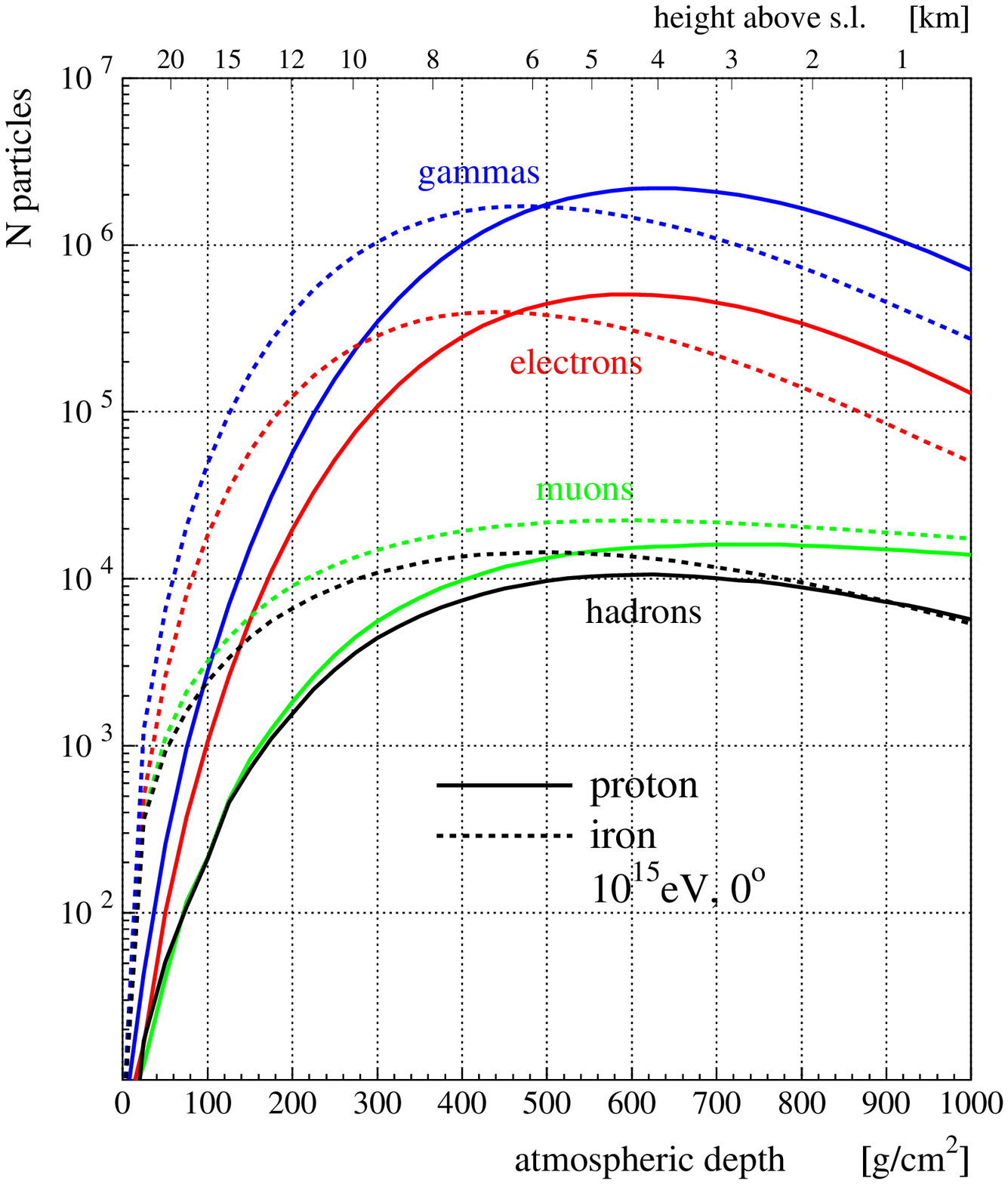,clip=,width=11.5cm}}

\centerline{\em Fig.5 Longitudinal EAS development}

}

The most efficient observables with respect to the mass composition is the
correlation of total intensities of the electron and muon components 
(showers sizes N$_e$ and N$_\mu$). This is obvious from the inspection of the
different longitudinal development of the shower sizes (Fig.5).

For the comparison of the observables with the pseudo data we have to
realise: None of the observables is strictly only dependent on the mass of
primary, or only dependent from the energy, and since we are investigating
an a-priori unknown spectral distribution accompanied by another a-priori
unknown variation of the elemental composition (or vice versa), there is
always an intriguing feedback of the estimates of both. Therefore
multivariate analyses, correlating the observations of different EAS
variables are strongly required, and the inference from only one EAS
component has been often misleading. For the analysis of the correlated
distributions without any bias of a constraining parameterisation, there are
adequate methods worked out involving neural networks and Bayesian decision
making [10,11]. Applying these techniques, for each particular case, i.e. for
a particular set of selected EAS variables or for a chosen number of mass
groups or for a specific hadronic interaction model generating the reference
patterns, matrices for true - and misclassification are obtained. From
that measures for the confidence and errors can be constructed.

\section{The Knee}

It is currently believed that cosmic rays are accelerated in a process called
diffusive shock acceleration. Suitable astrophysical shocks occur in
supernova explosions and the particles of the interstellar medium gain
energy as they are repeatedly overtaken by the expanding shock wave. Such a
mechanism leads in fact to a power law spectrum with the maximum energy of
about 
\mbox{Z$\cdot 10^{14}$\,eV} [12].
The upper limit E$_{max} \propto Z (\mbox{\bf r} \times \mbox{\bf B})$ reflect
the dependence from the size and the magnetic field of the accelerator
region. Alternatively the knee has been qualitatively explained by the leaky
box model that the galactic magnetic field let escape first the protons due
to their larger stiffness at the same energy compared to Fe.
In order to constrain the models and conjectures a
better knowledge of the shape of energy spectrum around the knee is quite
important. In particular, all approaches accounting for the origin and
acceleration mechanism, imply specific variation of the elemental
composition of primary cosmic rays, sometimes in a very detailed manner.
That are the issues addressed by the KASCADE experiment set up in
Forschungszentrum~Karlsruhe.

The concept of the KASCADE experiment with a multi-component detector array
is to measure a larger number of EAS variables for each individual event
with high accuracy. For this aim the detector has been designed. Specific
EAS variables accessible, in addition to the shower size N$_e$ and the
truncated muon number N$_\mu^{tr}$, are the number of hadrons N$_h^{100}$ with energies
larger than 100 GeV, the energy sum$\sum \mbox{E}_h$ of these hadrons, the energy
of the most energetic hadrons E$_h^{max}$, the number N$^*_\mu$ of muons with energies
larger than 2 GeV and others like some quantities representing the muon
arrival time distribution.

\vspace{-.2em}

\vbox{
\centerline{\epsfig{file=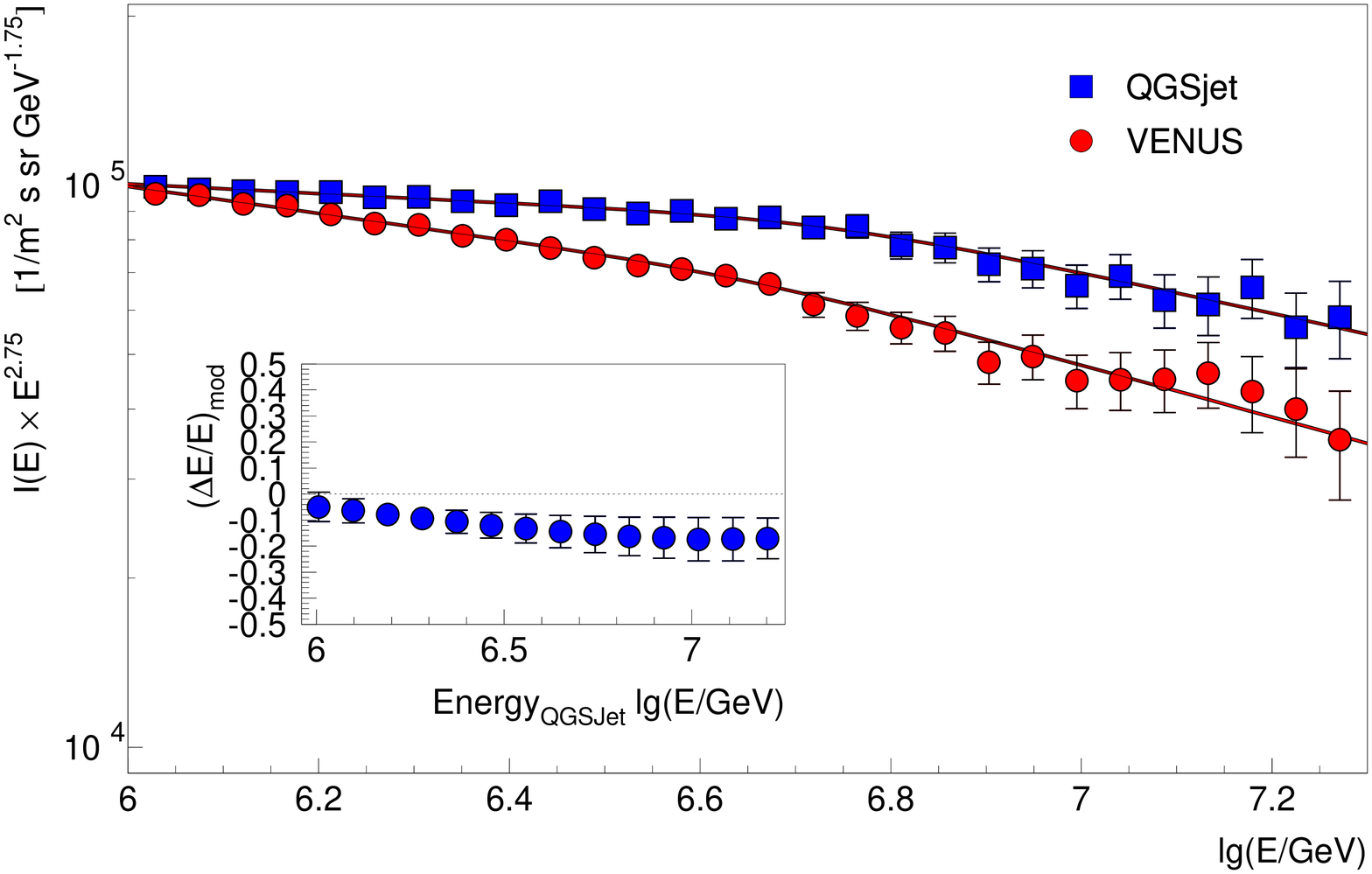,clip=,width=12cm}}

\vspace{-.1em}

\setbox0=\hbox{\em Fi g.6}
{\em\hangafter1\hangindent\wd0 \noindent Fig.6 The primary energy spectrum
  around the knee resulting from KASCADE data analyses on basis of two
  different hadronic interaction models [13,11].

}

}

\vspace{0.3em}

Fig.6 presents two solutions with the same data and the same analysis
procedures, but based on reference patterns from different hadronic
interaction models. This indicates the present limits due to the unavoidable
model dependence of any analysis.

Furthermore the result (we have much confidence in the QGSJet result) should
be seen under various aspects of current controversial discussions: Is there
an abrupt break in the spectrum (how the Akeno observations claim) and
where it is located? Or is the change of the spectral index rather smooth
as seemingly observed in the high altitude Tibet array (4300 m a.s.l.)?

Some years ago a hypothesis about origin of cosmic rays around the knee, a
theory propagated by Erlykin and Wolfendale [14], predicted a modulation in
the energy spectrum, wiggles due to the various mass production spectra of a
single supernova explosion, localised only few hundred light years away from
our solar system. Our data do not support this conjecture.

What concerns the mass composition, in the moment we may characterise the
situation by the energy spectra of various mass groups (Fig.7) resulting from a
non-parametric analysis i.e. by the most unbiased KASCADE result [13,11]
with the feature: The knee is made by the proton component only, and with
the question: Where is the iron knee? That is the focus of KASCADE GRANDE [8].

{
\def\piccom#1#2#3{\ifvmode\else{\parskip0pt\par\nointerlineskip}\fi\vbox to
  0pt{\kern-#2\hbox{\kern#1#3}\vss}\nointerlineskip}

\vbox{
\centerline{\epsfig{file=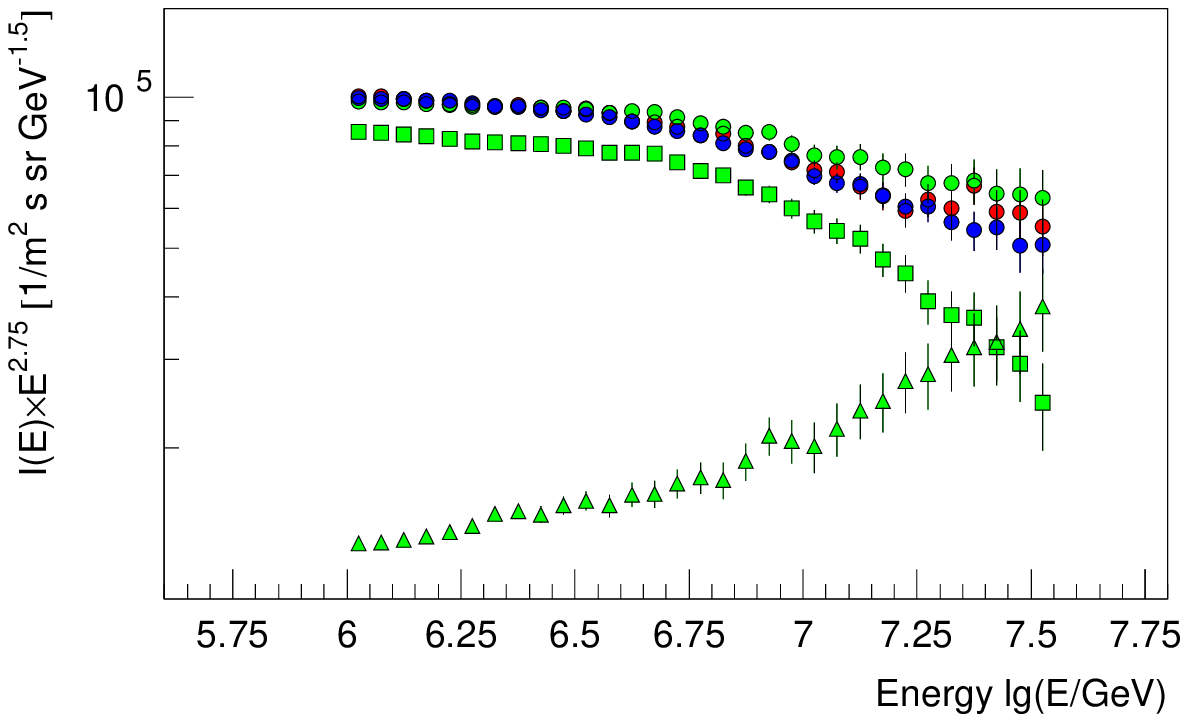,clip=,width=12cm}}
\piccom{4cm}{2.4cm}{\small\sf heavy}
\piccom{4cm}{4.9cm}{\small\sf light}
\piccom{4cm}{6cm}{\small\sf all}
\vspace{-.5em}
\centerline{\em Fig.7 Energy spectra of various mass groups [13]}

}

}

Figs. 6 and 7 display the present messages from KASCADE, analysed various
times with samples of large statistics and also with different methods, in
addition to nonparametric methods also with efficient parameterisations [15].

\vbox{
\centerline{\epsfig{file=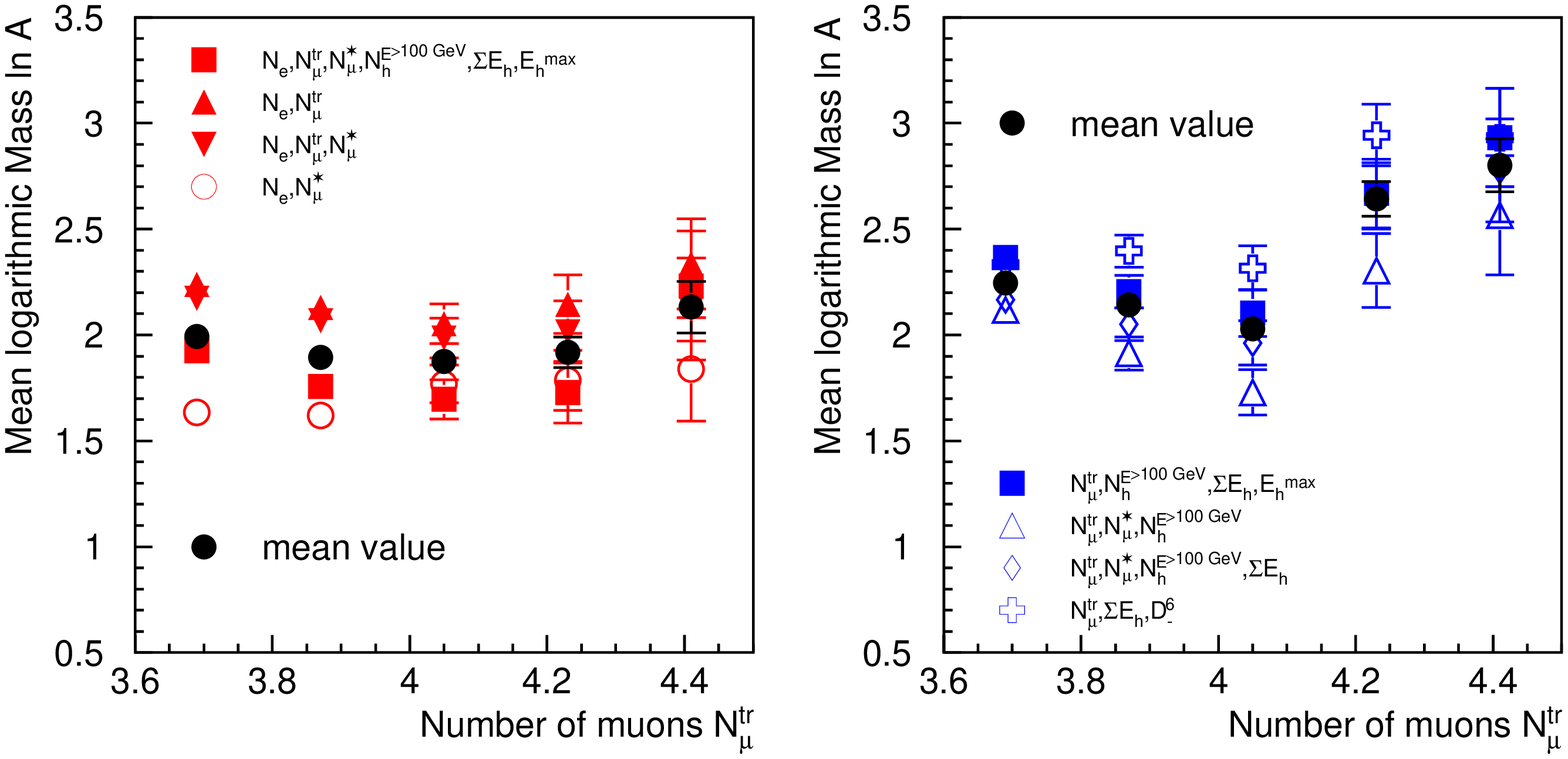,clip=,width=12cm}}

\setbox0=\hbox{\em Fi g.8} 
{\em\hangafter1\hangindent\wd0 \noindent
Fig.8 Variation of the mass composition $\langle ln A\rangle$
  inferred from KASCADE data analyzing various combinations of EAS
  observables [13].

}

}

\vspace{0.3em}

Fig.8 displays detail examples of the reconstructed chemical composition
represented by
the mean mass vs. energy identifier N$_\mu^{tr}$ (the knee is at $\log
\mbox{N}_\mu^{tr} = 4.1$) taking into account different sets of EAS
observables.
Compared are the results based on the hadronic interaction model
QGSJet [16]. We recognise the tendency that the lighter composition before
the knee get heavier beyond. The QGSJet model leads generally to a heavier
composition as compared to result of the VENUS [17] model. The reconstructed
mean mass depends obviously also from the correlation taken into account
This result of a feasibility study, implying a test of the used
interaction model, points to the way, how the data can be consistently
analysed on event -by - event basis with explorations of the particular
sensitivities and uncertainties, from the model dependence, e.g.

The focal points of studies around the knee can be summarised by the items:
\begin{itemize}
\em
\item The detailed shape of the energy spectrum

Smooth, with modulations or a sharp knee?

Variation of the elemental composition

Where is the knee of the iron component? Does it scale with Z or A?

\item Test of models of the production mechanisms

\item The hadronic interaction in the "terra incognita": $>10^{15}$\,eV

\end{itemize}

I emphasise again the basic dilemma of the present status. The analyses
of the measured data lead to results distinctly dependent from the
particular adopted high-energy interaction model. Though-in contrast to
other current experiments-the KASCADE experiment is able to specify the
inherent model dependence, thanks to the large number of observables,
studied simultaneously event per event, any progress needs an improved
knowledge of the interaction model.

\section{The Ankle and above}

In the range of the highest energies the first remarkable feature, in fact
establishing an enigma, stems from the existence of radiation fields which
fill the Universe. The 2.7\,K microwave background is the best known. Above
some thresholds the cosmic ray particles coming from long distances
inelastically interact with those background photons. High-energy incident
protons for which the background is blue shifted, start photo-pion
production above a few tens of EeV and get cooled down in this way: p +
$\gamma$\,(3K)$\to$ $\Delta (1232)$ $\to$ p + $\pi^0$ (n + $\pi^+$). That
is the predicted Greisen-Zatsepin-Kuzmin (GZK) spectral cut off [3]. The
consequence of the interaction with the radiation fields is that above
$5 \cdot 10^{19}$\,eV, photons, protons and nuclei have rather short attenuation
lengths, in the order of, say several tens Mpc, and the Universe gets
relatively opaque for them. To state this more explicitly: It is impossible
for ultrahigh energy cosmic particles to reach us from sources whose
distance would exceed 100 Mpc (this is roughly the size of our local
supercluster), unless rather exotic particles or interaction mechanisms are
envisaged.

A second feature is related to the chemical composition of ultrahigh energy
primary cosmic rays. If the highest energy cosmic rays would be mainly
protons, as some experimental results are tentatively interpreted, the
trajectories of single charged ultrahigh energy particles through the
galactic and extra-galactic magnetic fields (which are believed to be in
the order of $\mu$G and nG, respectively) get no more noticeably deflected over
distances limited by the Greisen-Zatsepin-Kuzmin cut-off. Typically the
angular deviation of a $10^{20}$\,eV proton from a source of 30~Mpc distance would
be about 2 degrees. In other words, above the cut-off, the direction of
incidence of such particles should roughly point to the source: Proton
astronomy should become possible to some extent, defined within the box of the
consequences of the cut-off. However, looking in our astrophysical
surroundings, the number of objects within a
distance of a few Mpc is quite limited, if such objects are even able to
accelerate particles to such extremely high energies at all.

What is the experimental knowledge?

The data around the ankle and above come from a few large-aperture ground
based detector arrays with two types of techniques (Tab.1).

\vspace{.5em}
\noindent
\begin{tabular}[c]{l@{\hspace{0.1em}}llrl}
\hline
\multicolumn{2}{l}{Array} & Location & Area & Principal Detectors \\
\hline
\vrule depth 0pt height 12pt width 0pt Haverah Park&[18] & England & 11\,km$^2$ & Water Cerenkov tanks \\[0.5ex]
Yakutsk&[19] & Russia & 10\,km$^2$ & Scintillation counters \\
& & & & Atmospheric Cerenkov \\[-.1ex]
& & & & detectors \\
& & & & Muon detectors \\[.5ex]
SUGAR&[20] & Australia & 60\,km$^2$ & Muon detectors \\[.5ex]
AGASA&[21] & Japan & 100\,km$^2$ & Scintillation counters \\
(Akeno)& & & & Muon detectors \\[.5ex]
Volcano Ranch&[22] & New Mexico & 8\,km$^2$ & Scintillation counters
\\
& & USA & & \\[.5ex]
Fly's Eye&[23] & Utah (USA) & & Air fluorescence detector \\[.5ex]
HiRes&[24] & Utah (USA) & & Air fluorescence detector \\
\hline
\end{tabular}

\vspace{.5em}

\centerline{\em Tab.1 UHECR detectors}

\vspace{.5em}

From historical reasons the smaller Volcano Ranch array is added because
there the first air shower event with the symbolic limit of $10^{20}$\,eV has been
observed [22]. Alternatively to particle detector arrays a second technique
is based on the observation of the nitrogen fluorescence induced by the
ionising particles crossing the air.

\vbox{
\centerline{\epsfig{file=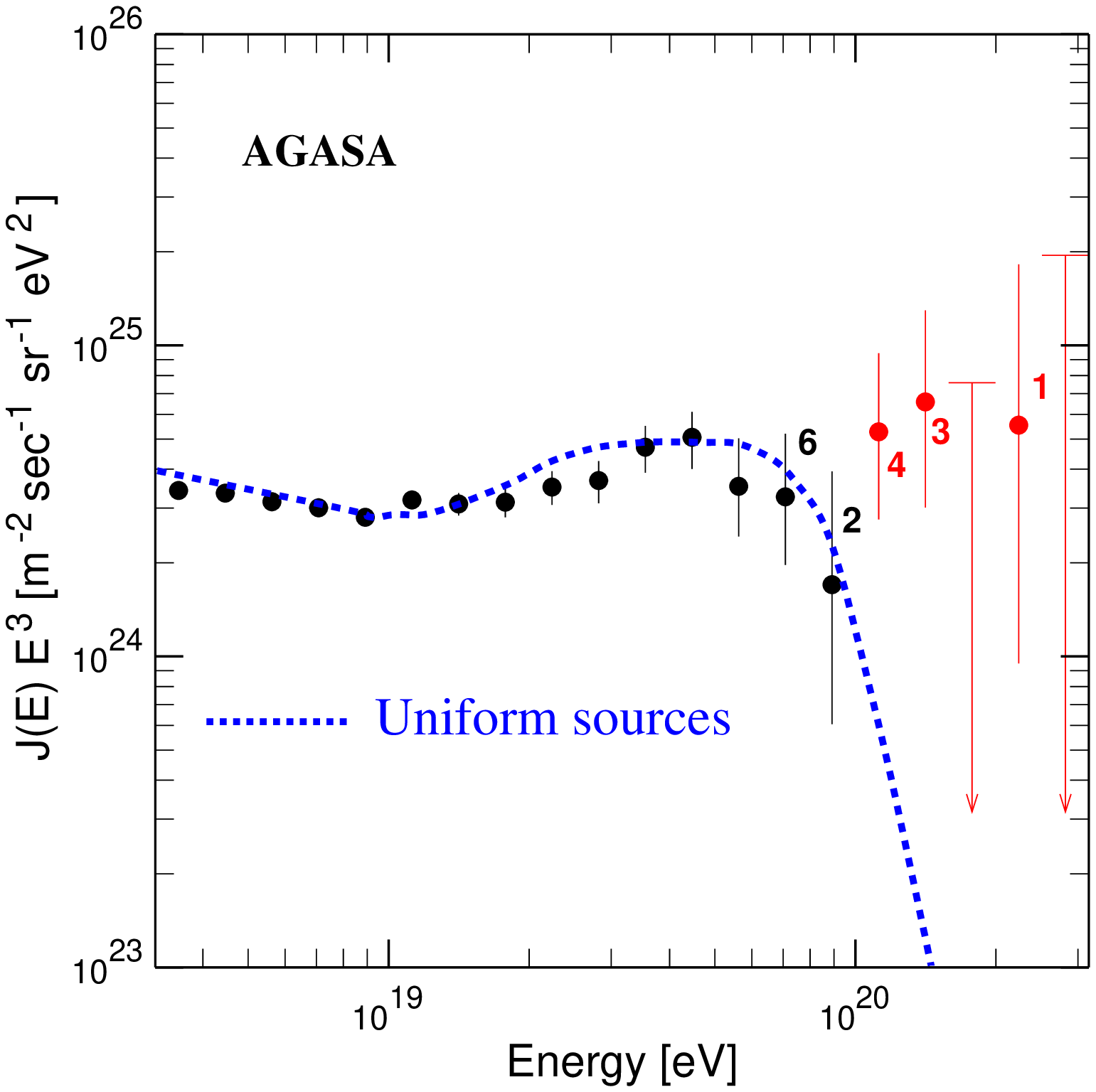,clip=,width=12cm}}

\centerline{\em Fig.9 Highest energy region of the cosmic ray spectrum [4]}

}
Fig.9 displays the highest energy region of the cosmic ray spectrum as
observed by the AGASA detector [4]. The figures near the data points
indicate the number of events and the bars show the 90\% confidence level.
The energy spectrum is multiplied by E$^3$, so that the part below $10^{18}$\,eV
becomes flat. The ankle structure becomes evident and the deviation from
the cut-off predictions. There are of course large error bars, but the
tendency is confirmed when other 13 events are included, detected by other
detectors (HiRes). The statistical accuracy of the distribution in the
supergalactic plane is to low to deduce any tendency.

The UHECR events constitute an enigma, when we ask: Where are the sites and
what are the acceleration mechanisms being capable to impart energies of
macroscopic orders (in the most energetic case of $3 \cdot 10^{20}$\,eV
equivalent to 50 joules) to a microscopic particle. Many processes have been
proposed, where in a astrophysical plasma large scale macroscopic motion is
transferred to individual particles, for example in a turbulence and by
shock waves. The crucial role plays the size of the acceleration region and
the magnetic field embedded in the plasma and keeping the gyroradius of the
particle in the acceleration region. That depends also from the velocity ß
of the motion. Under these aspects possible accelerator sites have
scrutinised [25,26]. If all parameters related to the question are taken into
account, one has to admit that none of the proposed scenarios seems fully
convincing. In addition we have to keep in mind that the sources should be
nearby in cosmological scales. Within the present statistical accuracy the
data do also not show a distinct correlation with nearby point sources.

However, if future studies would exclude "conventional" astrophysical
acceleration mechanisms, one would need to consider another class
of theories proposed as possible explanation, so-called "top-down"
processes (see ref. 27). Most of those study the possibility that UHECR
arise from decay of some super-heavy X particle whose mass is in the Grand
Unification range (10$^{25}$\,eV) produced during some phase transition period
during the early Universe. The models differ mainly, how to produce the
density of X particle to fit the UHECR observations and their survival
since some $10^{-35}$\,s after Big Bang. One should mention that such models and
conjectures have quite specific features and experimental
signatures (spectrum and mass composition) so that a discrimination appears
to be not impossible, provided the experimental knowledge could get
increased. That is just our challenge for the next generation of detectors
with large apertures!

\section{The next and over-next generation of detectors}

The next detector is the Pierre Auger observatory with 14.000\,km$^2$sr
aperture over two sites, one in each hemisphere [28].

The installation of the southern observatory (Fig.10) has started in 2000 with a
prototype array of 55 km$^2$ and a air fluorescence telescope, near the small
town of Malarg\"ue in the province of Mendozza, Argentina. Finally the site
will be equipped with 1600 detector stations (12\,m$^3$ tanks filled with
water detecting Cerenkov light produced by secondary particles),
distributed in a grid with 1.5\,km spacing. Four "eyes" composed of 30 air
fluorescence telescopes will view 3000\,km$^2$ of the site and measure
during clear moonless nights i.e. with a duty cycle of 10\% the giant
showers through the fluorescence generated in air. By this hybrid detector a
subsample of 10\% of the total number of events, simultaneously observed with
both techniques, enables a cross calibration and yield an unprecetended
quality for shower identification. It is expected to detect some 50 to 100
events per year above $10^{20}$\,eV, and 100 times more above $10^{19}$\,eV.

\vbox{
\epsfig{file=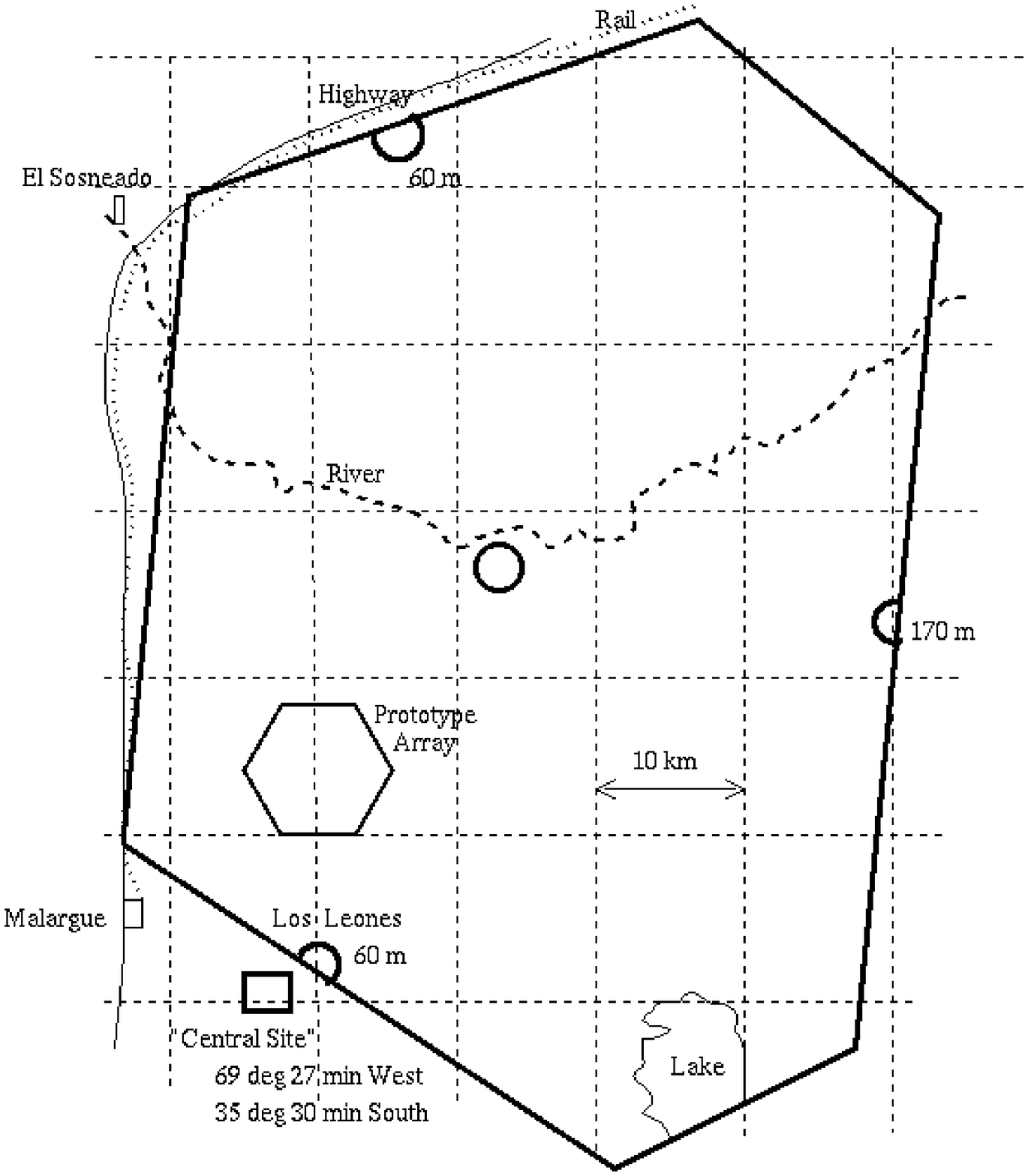,clip=,width=12cm}

\vspace{0.3em}

\centerline{\em Fig. 10 Layout of the southern Pierre Auger observatory [28]}

}

The Pierre Auger Project has just started, and the community looks already
forward to the next generation of detectors. There is less doubt that this
will be an air-borne detector observing the giant shower development in the
atmosphere with a huge aperture quasi " from above". This is envisaged with
the EUSO or ORBITING WIDE ANGLE LIGHT detector within the Airwatch project [29] 
by fluorescence detectors on satellites. This is particularly interesting when in few years
it will be shown that the spectral cut-off exceeds the reach the
Pierre Auger observatory and larger statistics is necessary for studies of
the focal points of Extremely High Energy Cosmic Ray Observations.

\noindent The questions are:

\begin{itemize}
\em
\item 
What is the reason for the change of the spectral index at the
"ankle"?

A change of the production mechanism?
A change of the elemental composition?
Or a change of the character of the interaction?

\item
How is the shape of the spectrum at energies above $10^{20}$\,eV and is
there a limit of maximum energy?

\item
Is there any directional correlation pointing to the sources of
UHCR?

\end{itemize}

With experimental answers to these questions we may provide some hints for
explaining the origin. However, there is the lesson of the advanced studies
of the knee region like with KASCADE, that the investigation of the
far-reaching astrophysical aspects by EAS observations has to be accompanied
by a serious and quantitative understanding of the hadronic interactions in
that energy range. That is the other side of the medal of necessary efforts!
Without that, even the energy determination of EAS and scale of the
spectrum may remain under debate! This debate got recently some new impact
since the HiRes collaboration presented a new calibration inducing some
doubt on the non-existence of Greisen-Zatsepin cut-off.

Let me conclude with the following remark. The most remarkable feature of the cosmic radiation 
is that the investigators have not yet found a natural end of the energy spectrum. We do not know 
the source of such a radiation, and the features establish a mystery of great cosmological 
relevance.

{\em
\noindent 
I would like to thank Dr.~Andreas~Haungs for clarifying discussions and
Dipl.Phys. Joachim~Scholz for technical help in preparing the script. 
I use the opportunity to express my sincere thanks to the organisers
of this meeting for providing me the possibility to participate.

}

\section*{References}

{
\def\nre#1{\item[\hbox to 0.8cm{\hfill[#1]}]}

\begin{list}{}{\itemsep-.1em}

\sloppy

\nre{1} V.Hess, Phys.Z.13(1912)1084

\nre{2} G.V.Kulikov and G.B.Khristiansen, Soviet Physics JETP 35(1959)441

\nre{3} K.Greisen, Phys.Rev.Lett.16(1966)748;\\ G.T.Zatsepin and V.A.Kuzmin,
Soviet Physics JETP Lett. 3(1966)78

\nre{4} M.Takeda et al.-AGASA collaboration, Phys.Rev.Lett.81(1998)1163

\nre{5} P.Auger et al., Comptes rendus, Académie des Sciences 206(1938)172;
ibid
207(1938)228

\nre{6} W.Kohlh\"orster et al., Naturwissenschaften 26(1938)576

\nre{7} H.O.Klages et al.-KASCADE collaboration,\\ Nucl.Phys.B
(Proc.Suppl.) 52B(1997)92

\nre{8} M.Bertaina et al.-KASCADE Grande collaboration,\\ Proc.27th ICRC Hamburg
2001,
Germany, p.796

\nre{9} D.Heck et al., FZKA Report 6019 (1998)\\ Forschungszentrum Karlsruhe

\nre{10} A.A.Chilingarian, Comp.Phys.Comm.54(1989)381;\\ A.A.Chilingarian and
G.Z.Zasian,
Nuovo Cim.14(1991)355

\nre{11} M.Roth,FZKA-Report 6262 (1999) Forschungszentrum Karlsruhe

\nre{12} P.O.Lagage and C.J.Cesarsky,\\ Astron.\&Astrophys. 118(1983)223 and
125(1983)249

\nre{13} T.Antoni et al.-KASCADE collaboration:\\ "A non-parametric approach to
infer the
energy spectrum and mass composition of cosmic rays", Astropart.
Phys.(2001) in press\\
M.Roth et al.-KASCADE collaboration,\\ Proc.27th ICRC Hamburg 2001,
Germany,
p.88

\nre{14} A.D.Erlykin and A.W.Wolfendale, Journ.Phys.G 23(1997)979

\nre{15} H.Ulrich et al.-KASCADE collaboration,\\ Proc.27th ICRC Hamburg 2001,
Germany,
p.97

\nre{16} N.N.Kalmykov, S.S.Ostapchenko and A.I.Pavlov,\\ Nucl.Phys.B (Proc.Suppl)
52B
(1997)17

\nre{17} K.Werner, Phys.Rep.232(1993)87

\nre{18} M.A.Lawrence, R.J.O. Reid and A.A.Watson, J.Phys.G 17(1991)773

\nre{19} B.N.Afanaviev et al., Proc. 24th ICRC 1995, Rome, Italy 2,756

\nre{20} L.Horton et al., Proc.20th ICRC 1987, Moscow, vol.1, p.404

\nre{21} S.Yoshida et al., Astropart.Phys.3 (1995)105

\nre{22} J.Linsley, Phys.Rev.Lett.10(1963)146

\nre{23} D.J.Bird et al., ApJ 424(1994)491

\nre{24} P.Sokolsky, Proc. AIP Conf., 1998, p.65

\nre{25} A.M.Hillas, Ann.Rev.Astron.Astrophysics 22(1984)425

\nre{26} S.Yoshida and H. Dai, J.Phys.G 24(1998)905

\nre{27} G.Sigl, S.Lee, P.Bhattacharjee and S.Yoshida,\\ Phys. Rev. D59(1999)116008

\nre{28} The Pierre Auger Project Design Report, Fermilab, October 1995
(www.auger.org)

\nre{29} O.Catalano et al., Proc.19th Texas Symposium on\\ Relativistic
Astrophysics, Paris, France,
1998

\end{list}

}

\end{document}